\documentclass[fleqn,12pt,twoside]{article}
\usepackage{espcrc1}
\usepackage{graphicx}

\title{Possible Hadronic Molecule $\Lambda$(1405) and 
Thermal Glueballs in SU(3) Lattice QCD}

\author{
H. Suganuma\address[titech]{Faculty of Science, Tokyo Institute of Technology, Tokyo 152-8552, Japan
\vspace{-0.3cm}}, 
N. Ishii\address[titech]{The Institute of Physical and Chemical Research (RIKEN), Wako 351-0198, Japan
\vspace{-0.3cm}},
H. Matsufuru\address[titech]{Yukawa Institute for Theoretical Physics, Kyoto University, Kyoto 606-8502, Japan
\vspace{-0.3cm}}, 
Y. Nemoto\address[titech]{Brookhaven National Laboratory, RBRC, New York 11973-5000, USA
\vspace{-0.3cm}} and 
T. T. Takahashi\address[titech]{RCNP, Osaka University, Mihogaoka 10-1, Osaka 567-0047, Japan
\vspace{-0.1cm}
}
}
       
\begin{document}

\maketitle

\begin{abstract}
\noindent
{\bf Abstract.}
{\small
We aim to construct quark hadron physics based on QCD. 
First, using lattice QCD, we study mass spectra of positive-parity and negative-parity baryons in the octet, the decuplet and the singlet representations of the SU(3) flavor. In particular, we consider the lightest negative-parity baryon, the $\Lambda$(1405), 
which can be an exotic hadron as the $N \bar K$ molecular state or the flavor-singlet three-quark state. 
We investigate the negative-parity flavor-singlet three-quark state in lattice QCD using the quenched approximation, 
where the dynamical quark-anitiquark pair creation is absent 
and no mixing occurs between the three-quark and the five-quark states. 
Our lattice QCD analysis suggests that the flavor-singlet three-quark state is so heavy that 
the $\Lambda$(1405) cannot be identified as the three-quark state, 
which supports the possibility of the molecular-state picture of the $\Lambda$(1405).
Second, we study thermal properties of the scalar glueball in an anisotropic lattice QCD, and find about 300 MeV mass reduction near the QCD critical temperature from the pole-mass analysis. 
Finally, we study the three-quark potential, which is responsible to the baryon properties.
The detailed lattice QCD analysis for the 3Q potential indicates the Y-type flux-tube formation linking the three quarks.
}
\end{abstract}

\section{Lattice QCD study for $\Lambda$(1405)}

Among a lot of hadrons, the $\Lambda$(1405) is a very special interesting hadron.
In spite of the strange baryon, the $\Lambda$(1405) is the lightest negative-parity baryon. In fact, the $\Lambda$(1405) is much lighter than 
the low-lying non-strange negative-parity baryons, the $N$(1520) with $J^{P}=\frac{3}{2}^{-}$ and the $N$(1535) with $J^{P}=\frac12^{-}$.
Moreover, there are two interesting physical interpretations on the $\Lambda$(1405).
\begin{itemize}
\item 
In the quark-model framework, the $\Lambda$(1405) is described as the flavor-singlet three-quark system. 
(cf. The H-dibaryon is also a flavor-singlet candidate.)
\vspace{-0.2cm}
\item
As another interpretation, the $\Lambda$(1405) is an interesting candidate of the hadronic molecule such as the $N\bar K$ bound state with a large binding energy about 30MeV. (cf. For deutrons, the binding energy is about 2.2MeV.)
\end{itemize}
In the valence picture, the $N\bar K$ hadronic molecule is described as $qqq$-$q\bar q$, and hence we call this state as the ``5-quark (5Q) state" for the simple notation.
Of course, in the real world, the 3Q state and the 5Q state would be mixed as 
\begin{eqnarray}
|\Lambda(1405)\rangle \simeq C_{\rm 3Q}|{\rm 3Q}\rangle+C_{\rm 5Q}|{\rm 3Q-Q}\bar {\rm Q}\rangle
\end{eqnarray}
So, the more realistic question is as follows. Which is the dominant component 
of the $\Lambda$(1405), the 3Q state or the 5Q state ?
The answer to this question would be important for the argument of the hyper-nuclei and some related subjects in astrophysics.
So, we investigate the $\Lambda$(1405) using lattice QCD, 
which is the first-principle calculation of the strong interaction.
However, also in lattice QCD, the 3Q and the 5Q states are mixed automatically through the $q$-$\bar q$ pair creation, and therefore 
it is almost impossible to distinguish the 3Q and the 5Q states in lattice QCD.

To overcome this difficulty, we use the advantage of the quenched approximation, 
which does not include the dynamical $q$-$\bar q$ pair creation.
Then, in quenched QCD, we can investigate the 3Q and 5Q states, individually.
Note here that quenched QCD reproduces various hadron masses, and is widely used in lattice QCD simulations.
In fact, once the 3Q state is prepared as the initial state, the system continues to be the 3Q state during the time evolution 
in quenched QCD, with keeping essence of QCD.
We report here the lattice QCD test for the $\Lambda$(1405) in terms of the 3Q flavor-singlet state\cite{NMNS01}.
Also, we investigate all possible low-lying baryons with positive-parity 
and negative-parity in the flavor-octet, decuplet and singlet representations.

We adopt SU(3)$_c$ lattice QCD using the improved action with the clover fermion on the three different lattices with $\beta$=5.75, 5.95 and 6.10. 
We use anisotropic lattice with finer temporal lattice spacing as $a_s=4a_t$.

Most of hadron masses of pseudo-scalar mesons, vector mesons, positive-parity baryons and negative-parity baryons  
are reproduced within about 10\% deviation, as shown in Table~1. 
However, in lattice QCD, the flavor-singlet 3Q state seems rather heavy as 
\begin{eqnarray}
M({\rm 3Q}, {\rm flavor \ \  singlet}, J^P=1/2^-) \simeq 1.7{\rm GeV}.
\end{eqnarray}
In other words, the $\Lambda$(1405) seems so light that it cannot be identified as the flavor-singlet 
3Q state, which supports the $N\bar K$ molecule picture for the $\Lambda$(1405).

For more definite conclusion, we need to investigate the 5Q state in lattice QCD at the quenched level, 
as well as the full lattice QCD calculation for the $\Lambda$(1405).

\vspace{0.3cm}
\begin{minipage}{0.5\textwidth}
\small
\begin{tabular}{lccc}
\hline\hline
   & $\beta=5.95$ & $\beta=6.10$ & Exp.\\
\hline
$\rho$  & 0.7965(53)& 0.8005(65) & 0.770\\
$K^*$           &   0.892 (fit)   &   0.892 (fit)    & 0.892\\
$\phi$          & 0.9913(53)& 0.9873(66) & 1.020\\
\hline
$N$             & 1.0781(81)& 1.1055(72) & 0.939\\
$\Lambda$       & 1.1825(75)& 1.2002(67) & 1.116\\
$\Sigma$        & 1.1982(76)& 1.2173(67) & 1.193\\
$\Xi$           & 1.3183(75)& 1.3291(67) & 1.318\\
\hline
$\Delta$        & 1.342(16) & 1.3685(17) & 1.232\\
$\Sigma^*$      & 1.440(14) & 1.4586(15) & 1.385\\
$\Xi^*$         & 1.538(12) & 1.5486(13) & 1.530\\
$\Omega$        & 1.635(11) & 1.6387(12) & 1.672\\
\hline\hline
\end{tabular}
\end{minipage}
\begin{minipage}{0.5\textwidth}
\small
\vspace{2mm}
\begin{tabular}{lcccc}
\hline\hline
   & $\beta=5.95$ & $\beta=6.10$ & Exp. \\ \hline
$N^{(-)}$       & 1.599(59) & 1.618(57) & 1.535\\
$\Sigma^{(-)}$  & 1.705(49) & 1.717(49) & 1.620\\
$\Xi^{(-)}$     & 1.810(40) & 1.816(42) & --\\
\hline
$\Lambda^{(-)}_{\rm octet}$  
                & 1.703(48) & 1.700(49) & 1.670\\
$\Lambda^{(-)}_{\rm singlet}$
                & 1.646(49) & 1.725(39) & 1.405\\
\hline
$\Delta^{(-)}$  & 1.877(73) & 1.833(80) & 1.700\\
$\Sigma^{*(-)}$ & 1.955(61) & 1.913(69) & --\\
$\Xi^{*(-)}$    & 2.032(50) & 1.994(57) & --\\
$\Omega^{(-)}$  & 2.109(39) & 2.074(46) & --\\
\hline
$\Lambda^{(+)}_{\rm singlet}$
                & 2.292(46) & 2.150(70) & --\\
\hline\hline
\end{tabular}
\label{tab:spectrum_phys2}
\end{minipage}

\begin{small}
\begin{center}
{\bf TABLE 1.\ \ }The lattice QCD result for various hadron masses in GeV.
\end{center}
\end{small}

\section{Scalar glueballs and their thermal properties}

\subsection{The ``Higgs particle" in QCD}
``Can you imagine a molecule of photons ? \cite{I82}''
Of course, there is no photonic molecule in QED, because of the absence of 
the self-interaction between photons.
In QCD, however, due to its nonabelian nature, there appears the self-interaction among gluons, 
and gluonic bound states, glueballs, are predicted to exist like ordinary mesons. (In fact, two or more gluons can 
make color-singlet objects like ${\rm Q}$-$\bar {\rm Q}$ mesons.)
Anyway, the glueball, which was theoretically predicted in QCD, is an exotic bound-state of gauge fields.

The glueballs are color-singlet objects without valence quarks, 
so that they are flavor-singlet and baryonless as $I=S=B=0$.
Lattice QCD simulations predict the masses of glueballs as
$M(J^{PC}=0^{++})=1500-1700{\rm MeV}$ and $M(J^{PC}=2^{++})=2000-2200{\rm MeV}$.
The experimental candidate of the lowest scalar glueball is considered as 
$f_0(1500)$, which is hard to be explained with the quark model.

In a simple construction, the scalar glueball operator is expressed as 
$\Phi(x)=G_{\mu\nu}^aG^{\mu\nu}_a$, with $G_{\mu\nu}^{a}$ being the 
field strength.
However, in the nonperturbative vacuum of QCD, this operator is known 
to be condensed like the Higgs scalar as 
$\frac{g^2}{32\pi}\langle G_{\mu\nu}^aG^{\mu\nu}_a\rangle \simeq 
(200{\rm MeV})^4$, which is called as the gluon condensate.
Like the Higgs particle, the physically observed field is expressed as the shifted operator, $\tilde \Phi(x) \equiv G_{\mu\nu}^aG^{\mu\nu}_a- 
\langle G_{\mu\nu}^aG^{\mu\nu}_a \rangle$.
Thus, the scalar glueball is directly connected with the vacuum structure of nonperturbative QCD, and, as a consequence, the scalar glueball is expected to be rather sensitive to the QCD vacuum.
For instance, we expect a large thermal effect on the scalar glueball \cite{IST95}, especially, near the critical temperature $T_c$ of the QCD phase transition, because the nonperturbative aspect of QCD would be largely reduced near $T_c$.

For comparison, we mention about the thermal mesons and thermal baryons in lattice QCD. In spite of many model predictions on the thermal properties of mesons, the recent lattice QCD Monte Carlo simulations indicate no significant thermal effect on the various mesons masses as 
$M_{\rm meson}(T \simeq T_c) \simeq M_{\rm meson}(T=0)$
from the direct measurement of the pole mass from the temporal correlation of the meson operators.
As for the thermal baryon, there is no pole mass measurement in lattice QCD, yet.

\subsection{Thermal Glueballs in lattice QCD}

In general, the pole mass is measured from the temporal correlation of the operator as 
$G(t)\equiv \langle \tilde \Phi(t)\tilde \Phi^\dagger(0)\rangle$.
If $t$ is large enough, the pole mass $m$ is obtained from the exponential decreasing 
as $G(t) \sim e^{-mt}$ in the Euclidean metric.
At high temperature, however, the temporal distance is limited as $0<t<1/T$ in the imaginary-time formalism. Then, it is rather difficult to perform the accurate measurement of the pole mass at high temperature.
(Here, do not confuse the pole mass with the screening mass, which is measured from the spatial correlation. The screening mass is not the physically observed particle mass.)
For instance, in most lattice QCD simulations at finite temperature, the number of the lattice points is $4-6$ in the temporal direction. In such cases, the independent information on the temporal correlator $G(t)$ is only $2-3$, because of the reflection symmetry in the temporal direction.
Then, it is almost impossible to measure the pole mass in these lattices.

To overcome this difficulty and to perform the accurate measurement of the thermal mass, we take the following prescriptions in lattice QCD.
\begin{itemize}
\item
To get detailed information of the temporal correlation 
$G(t)=\langle \tilde\Phi(t)\tilde \Phi^\dagger(0)\rangle$, we adopt the anisotropic lattice, where the temporal lattice spacing is much finer than the spatial one as 
$a_s=4a_t$.
\item
To reduce the statistical error, we use a large number of gauge configurations:  5,500-9,900 gauge configurations are used at each temperature.
(cf. In most lattice studies, about 100 gauge configurations are used.)
\item
For the accurate measurement of the lowest glueball mass, 
we carefully construct the appropriate operator which largely overlaps 
with the lowest glueball state, using the smearing method.
\end{itemize}
Here, the smearing method is a standard technique to reduce the 
excited-state components in the operator, without any change of physics. 
In fact, we can single out the lowest-state operator on lattice with the 
smearing method.

We calculate the temporal correlation of the scalar glueball in lattice QCD with $\beta=6.25$ and $20^3 \times N_t$ with 
$N_t$=33,34,35,36,37,38,40,43,45,50,72.
From the pole-mass analysis of the lattice QCD data, we observe about 300 MeV mass reduction of the lowest scalar glueball near the critical temperature of the QCD phase transition\cite{IMS02}, as shown in Fig.1. This result may suggest an observation 
of the large thermal effect on the scalar glueball in future RHIC experiments.

\begin{figure}
\begin{center}
\includegraphics[scale=0.5]{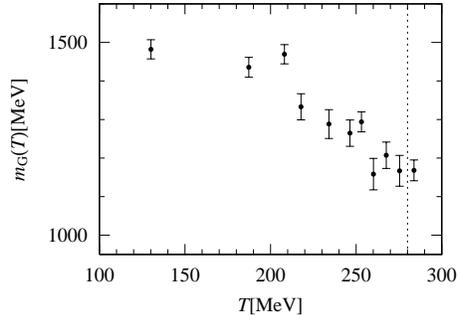}
\vspace{-0.5cm}
\caption{The lowest scalar-glueball mass $m_G(T)$ plotted against the temperature $T$. 
The vertical dotted line denotes the critical temperature $T_c \simeq 280{\rm MeV}$ in quenched QCD.}
\end{center}
\vspace{-0.5cm}
\end{figure}

\section{The Three-Quark Potential in SU(3) Lattice QCD}

In the modern science, to be based on the experiment and to find out the principle are two important viewpoints. To extract the principle or the essence of phenomena, some simplification is to be done, if necessary. Of course, such simplification does not mean to treat only simple system. Nevertheless, many theoretical particle physicists misunderstand this point.
The lattice QCD study of the three-quark baryonic potential was one of the typical blind spots in the elementary particle physics.

Using lattice QCD, the simple quark-antiquark (Q-$\bar {\rm Q}$) potential has been studied in detail. In contrast, there was almost no lattice QCD calculation for the three-quark (3Q) potential, although it plays the essential role to determine the baryon properties. Note here that the three-body force among the quarks is one of the ``primary force" reflecting the three colors in QCD. (cf. In most cases, three-body forces are regarded as residual interactions.) 

In fact, QCD has two kinds of inter-quark potentials, the Q-$\bar {\rm Q}$ potential and the 3Q potential, corresponding to mesons and baryons, respectively. 
The 3Q potential is also important for the study of the quark confinement, which is one of the most relevant features in QCD. For, one of the essential points of the quark confinement is the color-flux-tube formation and the appearance of the linear potential among quarks. The study of the 3Q potential tells us how to realize the quark confinement in baryons. Therefore, we study the 3Q potential in SU(3)$_c$ lattice QCD at the quenched level.

Now, let us consider the potential form in the Q-$\bar{\rm Q}$ and 3Q systems with respect to QCD. 
In the short-distance region, perturbative QCD is applicable and 
the Coulomb-type potential appears through the one-gluon-exchange (OGE) process.
In the long-distance distance at the quenched level, the flux-tube picture would be applicable from the argument of the 
strong-coupling expansion of QCD, and hence a linear confinement potential proportional to the total flux-tube length is expected to appear.
Indeed, lattice QCD results for the Q-$\bar {\rm Q}$ potential are well described by
$
V_{\rm Q \bar{Q}}(r)=-\frac{A_{\rm Q \bar{Q}}}{r}
+\sigma_{\rm Q \bar{Q}} r+C_{\rm Q \bar{Q}}
$
at the quenched level. In fact, $V_{\rm Q \bar{Q}}$ is described by a sum of the short-distance OGE result and the long-distance flux-tube result.

Based on the short-distance OGE and the long-distance flux-tube picture\cite{CI86}, 
we theoretically deduce the functional form of the 3Q potential $V_{\rm 3Q}$ as 
\begin{equation}
V_{\rm 3Q}=-A_{\rm 3Q}\sum_{i<j}\frac1{|{\bf r}_i-{\bf r}_j|}
+\sigma_{\rm 3Q} L_{\rm min}+C_{\rm 3Q},
\label{3Qpot}
\end{equation}
with $L_{\rm min}$ the minimal value of total length of flux tubes linking the three quarks: $L_{\rm min}={\rm AP}+{\rm BP}+{\rm CP}$ in Fig.2.

For more than 300 different patterns of the 3Q systems in total, 
we perform the accurate measurement of the 3Q potential $V_{\rm 3Q}$ 
using SU(3)$_c$ lattice QCD with $\beta$=5.7, 5.8 and 6.0 at the quenched level, 
and compare the lattice QCD data with the theoretical form of Eq.(\ref{3Qpot}) \cite{TMNS01,TSMN02}.
As a technical progress for the accurate measurement of $V_{\rm 3Q}$, 
we adopt the gauge-covariant smearing technique for the ground-state enhancement 
by removing many excited-state components such as flux-tube vibrational modes.

As a remarkable fact, the three-quark potential $V_{\rm 3Q}$ is well described by Eq.(\ref{3Qpot}) within about 1 \% deviation. 
(For detail, see Refs.\cite{TMNS01,TSMN02}.)
From the comparison with the Q-$\bar {\rm Q}$ potential, we find a universal feature of the string tension as $\sigma_{\rm 3Q} \simeq \sigma_{\rm Q \bar Q}$ and the one-gluon-exchange result for the Coulomb coefficient as $A_{\rm 3Q} \simeq \frac12 A_{\rm Q \bar Q}$.
In fact, the quark confinement in baryons is realized by the Y-type flux-tube formation, and the quark confining force is universal between mesons and baryons.

\begin{figure}
\begin{center}
\includegraphics[scale=0.8]{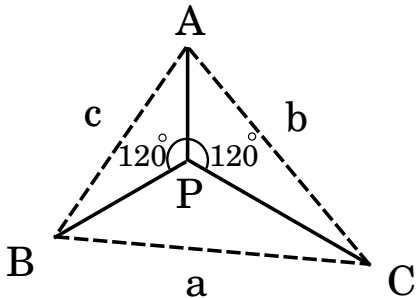}
\vspace{-0.5cm}
\caption{The flux-tube configuration of the 3Q system with the minimal value of the total flux-tube length. 
There appears a physical junction linking the three flux tubes at the Fermat point P.}
\end{center}
\vspace{-0.5cm}
\end{figure}

\section{Summary and Concluding Remarks}

We have studied the three topics of the quark hadron physics with SU(3)$_c$ lattice QCD.

\noindent
(1) We have studied negative-parity baryons with anisotropic lattice 
QCD. In particular, the $\Lambda$(1405), the lightest negative-parity baryon, 
is an interesting candidate of the molecular state of $N \bar K$ (the five-quark system) or the flavor-singlet three-quark state. 
We have analyzed the flavor-singlet three-quark state corresponding to the 
$\Lambda$(1405) in lattice QCD at the quenched level.
The point is to use the quenched approximation, where the dynamical 
quark-anitiquark pair creation is absent, and no mixing occurs between the three-quark and the five-quark states.
Our lattice QCD analysis suggests that the $\Lambda$(1405) is too light to be identified as a flavor-singlet three-quark state, 
which supports the $N \bar K$ molecular picture of the $\Lambda$(1405).

\noindent
(2) We have studied the thermal properties of the glueballs in SU(3)$_c$ anisotropic 
lattice QCD. From the pole-mass analysis, we have observed about 300 MeV mass reduction of the lowest scalar glueball near the critical temperature of the QCD phase transition. This result may indicate an observation of the large thermal effect on the scalar glueball in future RHIC experiments.

\noindent
(3) We have studied the static three-quark (3Q) potential for more than 300
different patterns of the 3Q systems in SU(3)$_c$ lattice QCD with $\beta$=5.7, 5.8 and 6.0. The lattice QCD 
data of the 3Q potential $V_{\rm 3Q}$ are well reproduced within 1 \% deviation by the sum of a constant, 
the two-body Coulomb term and the three-body linear confinement term proportional to the color-flux-tube length. 
From the comparison with the Q-$\bar {\rm Q}$ potential, we have found a universal feature of the string tension as 
$\sigma_{\rm 3Q} \simeq \sigma_{\rm Q \bar Q}$.
Thus, the detailed analysis of the 3Q potential indicates the Y-type flux-tube, which links the three quarks.


\begin{thebibliography}{99}
\bibitem{NMNS01}
N.~Nakajima, H.~Matsufuru, Y.~Nemoto and H.~Suganuma, \\
AIP Conf. Proc. {\bf CP594}, 349 (2001).
\bibitem{I82}
K. Ishikawa, Scientific American {\bf 247}, 142 (1982).
\bibitem{IST95} 
H.~Ichie, H.~Suganuma and H.~Toki,  Phys.~Rev.~{\bf D52},~2994~(1995).
\bibitem{IMS02}
N.~Ishii, H.~Suganuma and H.~Matsufuru, Phys.~Rev.~{\bf D66}, 014507 (2002).
\bibitem{CI86}
S.~Capstick and N.~Isgur, Phys.~Rev.~{\bf D34}, 2809 (1986).
\bibitem{TMNS01}
T.T.~Takahashi, H.~Matsufuru, Y.~Nemoto and H.~Suganuma, \\
Phys.~Rev.~Lett. {\bf 86},~18~(2001).
\bibitem{TSMN02}
T.~T.~Takahashi, H.~Suganuma, H.~Matsufuru and Y.~Nemoto, \\
Phys.~Rev.~{\bf D65}, 114509 (2002).
\end{thebibliography}
\end{document}